\documentclass[fleqn,usenatbib]{mnras}

\usepackage{newtxtext,newtxmath}

\usepackage[T1]{fontenc}
\usepackage{graphicx}	
\usepackage{amsmath}	
\usepackage{multicol}        
\usepackage{bm}		
\usepackage{pdflscape}	
\usepackage{hyperref} 

\usepackage{xcolor}

\usepackage{soul}

\newcommand{\diff}[2]{\frac{{\mathrm D} #1}{{\mathrm D} #2}}
\newcommand{\pder}[2]{\frac{\upartial #1}{\upartial #2}}

\newcommand{\vect}[1]{\mathbfit #1}
\newcommand{\grad}{ {\bf \nabla } }

\DeclareRobustCommand{\VAN}[3]{#2}
\let\VANthebibliography\thebibliography
\def\thebibliography{\DeclareRobustCommand{\VAN}[3]{##3}\VANthebibliography}

\usepackage{graphicx}	
\usepackage{amsmath}	

\title[Thermal misbalance \& acoustic-gravity waves]{Impact of thermal misbalance on acoustic-gravity waves in the solar atmosphere}

\author[D. S. Riashchikov et al.]{
D. S. Riashchikov,$^{1,2}$\thanks{E-mail:ryashchikovd@gmail.com}
N. E. Molevich,$^{1,2}$
and D. I. Zavershinskii$^{1,2}$
\\
$^{1}$Department of Theoretical Physics, Lebedev Physical Institute, Novo-Sadovaya st. 221, Samara, 443011, Russia\\
$^{2}$Department of Physics, Samara National Research University, Moscovskoe sh. 34, Samara, 443086, Russia\\
}

\date{Accepted XXX. Received YYY; in original form ZZZ}

\pubyear{2015}

\begin{document}
\label{firstpage}
\pagerange{\pageref{firstpage}--\pageref{lastpage}}
\maketitle

\begin{abstract}

The joint effect of gravity and thermal misbalance on the dynamics of acoustic-gravity waves (AGW) in the solar atmosphere is considered. It is shown that the heating and cooling taken in the form of power functions lead to the linear dependence of stationary temperature profile.  Estimates of the ratio of the characteristic length associated with thermal processes to the gravitational height show a predominant influence of thermal processes in the temperature range up to 2 MK and a comparable influence on the dynamics of AGW in the range from 2 to 10 MK. A study of the dispersion properties of AGW in an isothermal atmosphere showed that in regimes with an overwhelming influence of thermal processes, the acoustic cut-off frequency decreases up to $\sqrt{\gamma}$ times. At the same time, the maximum frequency of the gravitational mode (analog of the Brunt–Väisälä frequency in the medium without non-adiabatic heating and cooling) decreases with increasing power of thermal processes, and then the gravitational mode can become purely oscillatory.
\end{abstract}




\section{Introduction}
	\label{s:Introduction} 
Solar atmosphere is the natural plasma physics laboratory, which allows various oscillatory and wave motions to be observed and analyzed.  
One of the main advantages of such a laboratory is the number of wave modes for interpretation. These waves are oscillating over a wide range of periods and can be observed in the different regions of the solar atmosphere.
In particular, the 5-minute oscillations are routinely observed in the photosphere. Such oscillations are interpreted as global standing acoustic waves \citep{2005LRSP....2....6G}.
The chromosphere is dominated by the 3-min oscillations \citep{2012A&A...539A..23S} that can be explained by two main mechanisms: a chromospheric cavity (or resonator) \citep{2011ApJ...728...84B} and oscillations at the acoustic cut-off frequency \citep{1991A&A...250..235F}.
In the upper layers of the solar atmosphere, the observed wave periodicity generally varies from minutes to tens of minutes \citep{2019ApJ...874L...1N}. 
The density perturbations with periods of about an hour and several hours \citep{2014A&A...563A...8A,2015ApJ...807..176V} have been indicated in the upper corona and in the solar wind.

Another advantage of such a natural laboratory is the presence of a quite diverse set of processes affecting the wave dynamics and their dispersion properties. Some of the most important processes are magnetic field, gravity, and non-adiabatic processes (heating, cooling, and thermal conduction). Their characteristic spatial and temporal scales vary independently in the wide range of values. Due to this fact, not only one, but some combination of dispersion sources can affect magnetoacoustic (MA) waves.

The magnetic field is of special importance in the upper layers of the solar atmosphere. It provides the long-living existence of various magnetic structures like coronal loops, plumes, and prominences, which play the role of the waveguides for MA modes \citep[see, e.g.,][for recent reviews]{Nakariakov2020,2021SSRv..217...76B}. The wave-guiding dispersion connected with the field-aligned filamentation was originally investigated by \citet{1975IGAFS..37....3Z, 1982SvAL....8..132Z} and \citet{1983SoPh...88..179E}. Nowadays, the results of this theory are actively used for interpretation of the observational data. Furthermore, MHD-theory allows to use the slow \citep{2021SSRv..217...34W} and fast \citep{2021SSRv..217...73N,2020SSRv..216..136L} MA waves as a diagnostic tool for analysis the coronal plasma parameters.

The problem of gravitational stratification effect on dispersion properties of compressional waves was initially raised in the context of waves in the Earth's atmosphere \citep{1932hydr.book.....L}. It turns out that stratification introduces a very specific mode, namely, the gravity wave (GW), which has no possibility for vertical propagation in the atmosphere. As regards the astrophysical plasma, interest in GW was manifested in order to explain the uniform rotation of the solar core \citep{1997Natur.388..324G}.
These waves were also considered as a source of the mechanical heating of stellar atmospheres and corona \citep{1981ApJ...249..349M}. Due to the lack of direct observations, their contribution is usually neglected. However, analysis conducted by \citet{2008ApJ...681L.125S} reveals that the low-frequency atmospheric gravity waves are energetically more important for the lower solar atmosphere than high-frequency acoustic waves.

It is a well-known fact that gravitational stratification limits the spectrum of vertically propagating waves introducing so-called cut-off frequency for acoustic perturbations. It can also cause the spreading and appearance of the trailing oscillatory wake for broadband pulses. As we have mentioned previously, this effect is of great importance for the chromospheric oscillations \citep{1991A&A...250..235F,2012A&A...539A..23S}. Moreover, the cut-off effect is also actively studied in the context of perturbations in the upper solar corona and in the solar wind. The density structures with the periods lying in the range of 65–100 min and characteristic timescale of $\sim 90$ min has been indicated in the solar wind at 1 AU by \citet{2010SoPh..267..175V, 2015ApJ...807..176V}. The waves with the same characteristic period, detected at a distance of 2.5-15 solar radii, were indicated by \cite{2017A&A...601A..42P}. It has been proposed that standing shock at the 90 min acoustic cut-off period of the corona drives reconnection and induces these density perturbations observed in the solar wind \citep[see][for details]{2017A&A...601A..42P}. In the case of purely vertical propagation of acoustic modes in a medium with an arbitrary background temperature profile, the evolution of waves can be described by the Klein-Gordon equation. Such an approach was applied by \citet{2006RSPTA.364..447R} to describe the evolution of MA waves in the stratiﬁed atmosphere penetrated by a uniform vertical magnetic field. Furthermore, applying the similar approach, \citet{2015A&A...582A..57A}  showed with the help of the derived evolutionary equation that the cut-off period can vary with height and significantly decrease in the exponentially divergent magnetic flux tube with low-beta plasma. The

The solar atmosphere is also known as the non-adiabatic plasma. In such a medium, the waves are affected not only by dissipative processes like thermal conduction or viscosity, but also by temperature and density dependent radiative cooling \citep{Dere1997, Delzanna2020chianti} and heating processes \citep{1978ApJ...220..643R,2006A&A...460..573C}. In fact, the compression perturbation can disturb the balance of heating/cooling processes allowing 
the feedback between the plasma and the wave to take place. In the solar physics community, this feature is known as the so-called thermal misbalance \citep[see, e.g.,][]{10.1088/1361-6587/ac36a5, 2021SoPh..296..105P, 2022SoPh..297....5P, 2020SoPh..295..160B}. This feedback may affect the perturbation in a variety of ways including the dispersion of phase and group speeds \citep{Zavershinskii2019, doi:10.1121/2.0000505}, wave amplification or attenuation \citep{Kolotkov_2020}, additional phase shift between perturbations of various plasma parameters, etc. \citep[see, e.g.,][for details]{2021SoPh..296...96Z, 2021SoPh..296..105P, 2022SoPh..297....5P}. The constructed theory of thermal misbalance allows to introduce the relation between the parameters of waves (which can be observed) and coronal heating rate. Assuming the heating rate to be a power-law function of plasma density and temperature ($H(\rho,T) \sim  \rho^aT^b$), one may obtain some constraints on power indices $a$, $b$ and thereby narrow the set of possible forms of the heating function \citep[see, e.g.,][]{Kolotkov_2020, 2020SSRv..216..140V,2022MNRAS.514L..51K}. 

The combination of gravitational and magnetic effects on the evolution of compressional perturbations has been analyzed both analytically and numerically \citep[see, e.g.,][]{2019A&A...623A..62K,2019MNRAS.487.1489K}. In turn, the analysis of the combined effect of gravity and non-adiabatic processes is mostly limited to numerical models \citep{2020ApJ...896L...1M, 2021MNRAS.505...50G}. In our paper, we will try to contribute to solving the problem of the lack of analytical models and  make some predictions, which we believe can find their application in the interpretation of both observational data and numerical modelling results.

Our paper is organized in the following way. In Section~\ref{s:Model}, we discuss the basic equations and used assumptions. Further, in Section~\ref{s:Stationary state}, we describe the stationary state in the thermally active plasma and introduce possible height profiles of plasma parameters. Section~\ref{s:Isothermic case} is devoted to the waves in the isothermal gravitationally stratified plasma with the thermal misbalance. In Subsection~\ref{ss:Dispersion relation}, we introduce the dispersion relation for the AGW. Further, we show the temperature ranges, where the thermal misbalance impact on compression waves is comparable to or greater than the gravity effects (see Subsection~\ref{ss:The significance of thermal misbalance}). We describe how thermal misbalance affects the cut-off period and dispersion properties of AGW in Subsection~\ref{ss:Dispersion properties of acoustic-gravity}. Finally, the discussion and conclusions are presented in Section~\ref{s:Discussion}.

\section{Model}
	\label{s:Model} 
In our study, we will consider a fully-ionized gravitationally stratified plasma, and take into account the influence of heating and radiative cooling. 

One of the main focuses of our research is the analysis of slow MA waves guided along a magnetic field in coronal loops. The evolution of these modes in the general case can be described using the classic approach introduced by \cite{ 1982SvAL....8..132Z,1983SoPh...88..179E}. In the case of thin coronal tubes, one may apply second-order thin flux tube approximation \citep{Zhugzhda96}. The latter approach allows us to describe wave-guiding dispersion in a simpler way, without using hyperbolic and special functions. However, in the highly magnetized plasma, the wave-guiding dispersion prescribed by the magnetic structuring becomes sufficiently weak. In particular, well-known tube speed frequently associated with the slow-wave phase speed and applied for seismological needs  \citep{2007ApJ...656..598W, 2016NatPh..12..179J} is equal to sound speed under such conditions. In this case, the influence of the magnetic field on the dispersion properties of slow waves is rather weak, and their evolution can be described by the system of hydrodynamic equations with sufficiently high accuracy. Such an approach is known as infinite magnetic field approximation. It has demonstrated its effectiveness for the description of slow waves in coronal loops \citep{ 2000A&A...362.1151N, 2009A&A...494..339O, 2015ApJ...811L..13W, Zavershinskii2019, 2021SoPh..296..105P, 2022SoPh..297....5P} and also in stellar loops \citep{2018ApJ...856...51R, 2022ApJ...931...63L}. Subject to some restrictions on the radiative loss function \citep{1986ApJ...308..975A}, a similar system of initial equations can be applied to describe acoustic waves in the chromosphere \citep{2021ApJ...912...25A,2014ApJ...795...10L}. Our interest is the chromospheric regions with the plasma beta $\beta \gg 1$ and $\beta \ll 1$  (see calculated values of  plasma beta function at heights from 0.25 Mm to 2.5 Mm, shown in Figure 1 in  \cite{Bourdin_2017}  and Figure 3 in   \cite{2001SoPh..203...71G} ), where the influence of the magnetic field on the behavior of acoustic ($\beta \gg 1$) and slow ($\beta \ll 1$) waves can be neglected. In the calculations that will be presented in Section~\ref{ss:Dispersion properties of acoustic-gravity}, we will also look at photospheric parameters. We note, however, that these calculations will be of an illustrative nature. Since the assumption that radiation losses in the photosphere can be modelled by optically thin radiation is rather rough. The purpose of the calculations for this domain will be to show the importance of radiation transfer processes, in the context of acoustic wave evolution.

With the foregoing as background, the fundamental equations in the considered model are: 
\begin{equation}
    \pder{\rho}{t}+\grad\cdot\left(\rho{\vect{v}}\right)=0 \,,
	\label{eq:MassConservation}
\end{equation}
\begin{equation}
  \rho \diff{\vect{v}}{t} = -\grad P + \rho \vect{g},
	\label{eq:MotionConservation}
\end{equation}
\begin{equation}
    C_{V} \diff{T}{t} -\frac{k_\mathrm{B} T}{m \rho} \diff{\rho}{t}=-Q(\rho,T) + \frac{1}{\rho} \grad \left(\kappa \grad T \right),
	\label{eq:EnergyConservation}
\end{equation}
\begin{equation}
    P=\frac{k_\mathrm{B}}{m}\rho T .
	\label{eq:EOS}
\end{equation}

In Equations~(\ref{eq:MassConservation}) -- (\ref{eq:EOS}), $\rho$, $T$, and $P$ means the density, temperature, and pressure of the plasma, respectively. Vectors \vect{v} and \vect{g} are the plasma velocity and the gravitational acceleration, $\kappa$ is the thermal conduction coefficient. The coefficient $\kappa$ is assumed to be constant in the scope of this work. The Boltzmann constant and the mean mass per volume are respectively shown by $k_\mathrm{B}$ and $m$. In addition, $D/{D\textit{t}}=\upartial/\upartial t+\mathbfit{v}\cdot\nabla$ stands for the convective derivative.

In our analysis, we will use the Cartesian coordinate system \textit{x}, \textit{y}, \textit{z} and assume that the plasma is stratified along \textit{z}-axis (i.e.  gravity acceleration can be written as $\vect{g} = - g \vect{e_z}$, where $\vect{e_z}$ is the unity vector of $z$-axis).

The non-adiabatic processes in our model are described by the heat-loss function $Q\!\left(\rho, T\right)\!=\!L\!\left(\rho, T\right)-H\!\left(\rho, T\right)$, which is the difference between radiative cooling $L\!\left(\rho, T\right)$ and heating $H\!\left(\rho, T\right)$. The radiative losses   $L(\rho, T)$ in the optically thin plasma generally can be expressed as the power-law function:
\begin{equation}
L(\rho,T)= \psi \rho T^{\alpha},
\label{Cooling}
\end{equation}
where $\psi$ and $\alpha$ are piecewise constants depending on the temperature. The values of these constants can be calculated with the help of CHIANTI atomic database \citep{Delzanna2020chianti}. The use of optically thin radiative cooling is quite reasonable for coronal or almost coronal conditions (where the “coronal approximation” of  radiation loss rates as functions of temperature and density for a solar mixture of elements is applicable with sufficiently high accuracy \citep{1986ApJ...308..975A}). This approach can be also extended to modelling the processes in the upper chromosphere \citep{2012A&A...539A..39C, 1986ApJ...308..975A}. In turn, to model the photospheric radiation, one should solve the radiation transfer equation \citep{1984oup..book.....M}. 
However, some attempts are made to extend the power-law radiative cooling function to the lower temperature regions, such as 
partially ionised plasmas in prominences up to $T = 6000\mathrm{K}$ \citep[see e.g.][and the references therein]{2022SoPh..297..144I}. We also recommend an up-to-date review concerning the modelling of radiation processes in the solar atmosphere \citep{2020LRSP...17....3L}.

The heating  rate $H(\rho,T)$  is usually modeled by the power dependence on the thermodynamic parameters of the plasma \citep{Rosner1978,Dahlburg1988,Ibanez1993}:
\begin{equation}
H(\rho,T)=h \rho^aT^b,
\label{Heating}
\end{equation}
where $h$ is a constant calculated in order to balance cooling under steady state conditions $ (H(\rho_0,T_0)=L(\rho_0,T_0))$; $a$ and $b$ are constants determined by a specific heating mechanism.

Specific heating mechanisms (e.g., heating by coronal current dissipation, heating by Alfvén mode/mode conversion, etc.) written in the form of power functions of thermodynamic quantities were introduced by \cite{1978ApJ...220..643R}. These heating mechanisms expressed in the form of power functions of temperature and density with parameters $a$ and $b$ can be found in \cite{Ibanez1993} (the unit of heating function in this work is $\mathrm{ergs \; cm^{-3} \; s^{-1}}$, while in the current paper, it is $\mathrm{ergs \; g^{-1} \; s^{-1}}$). However, the study of \cite{Kolotkov_2020} has shown that the thermal mode in the considered heating mechanisms is always unstable in the case of suppressed thermal conduction and can be either stable or unstable in the case of finite conduction. The acoustic mode also can be stable or unstable. Thus, it was proposed to constrain the values of parameters $a$ and $b$ in order to ensure the stability of the solar atmosphere, and the calculated characteristics of acoustic waves (period, propagation velocity, decay time) corresponded to observational data \citep{Kolotkov_2020, 2022MNRAS.514L..51K}. Similar constraints obtained with various observational data and analytical models, taken together, will probably allow us to narrow the range of admissible values of $a$ and $b$ and to identify the possible mechanisms that cause the obtained dependence of the heating power on density and temperature.

In this paper, we propose an analytical model that relates the parameters of the heating function $a$ and $b$ to the height profiles of temperature, pressure, and density (see Section~\ref{s:Stationary state}).

\section{Stationary state}
	\label{s:Stationary state}
To analyze the properties of AGW, first, we should specify the stationary state in the  plasma under consideration. Due to the fact that the analyzed plasma is the non-adiabatic one, the thermal balance requires special attention.
 
The thermal balance implies that the right-hand side of Eq.~(\ref{eq:EnergyConservation}) should equal 0. 
In general, the  heating $H\!\left(\rho, T\right)$ and radiation cooling  $L\!\left(\rho, T\right)$  can be locally unbalanced ($Q(\rho,T) \neq 0$). In this case, the local excess of heat will be transferred to other regions by the thermal conduction in order for another steady state to occur. Similar mechanism together with thermal instabilities is considered to be a possible reason for the coronal rain  formation \citep{Antolin2020_TNE, 2022ApJ...931L..27S}. The temperature perturbation of a steady state resulting from a certain initiating event (e.g., growth of the compression perturbation caused by  thermal instability) in some part of the coronal loop propagates throughout the loop due to thermal conduction and gravitational forces. Moreover, under some conditions, the loop state can become continuously evolving even though the heating remains stationary. The plasma is constantly adjusting to imbalances in the energies and forces, essentially searching for a nonexistent equilibrium \citep{2019SoPh..294..173K}. This effect is known as thermal non-equilibrium or TNE cycle \citep{2022FrASS...920116A, 1991ApJ...378..372A}. In the current research, we assume that there is a stable state of equilibrium with heating and cooling balancing each other ($Q(\rho, T) = 0$) and there are no constant energy fluxes due to thermal conduction. This prevents effects such as coronal rain and TNE cycles from occurring and allows a stationary height profile of temperature, density, and pressure to exist.
 
Therefore, considering thermal balance, we  will assume that heating and cooling rates balance each other ($\!L\!\left(\rho_0, T_0\right) = H\!\left(\rho_0, T_0\right)$, or $Q\left(\rho_0, T_0\right)= 0 $) in the stationary state. We will also consider the cooling and heating rates in forms (\ref{Cooling}) and (\ref{Heating}), respectively. Thus, the local thermal balance can be written in the form shown below:
 \begin{equation}
    \psi \rho_0 T_0^{\alpha} = h \rho_0^a T_0^b.
	\label{eq:Thermalbalance}
\end{equation}

As will be seen later the heating and cooling taken in the form of power law functions together with the condition of local thermal equilibrium $Q\left(\rho_0, T_0\right)=0 $ lead to the {linear dependency of temperature with height}. In this case, the thermal conduction term vanishes and the condition of equality to zero of the right-hand side of Eq.~(\ref{eq:EnergyConservation}) is satisfied.

Considering Eq.~(\ref{eq:Thermalbalance}) and excluding temperature with the help of the equation of state (\ref{eq:EOS}), one can obtain the following relation between gas-dynamic pressure and density:
\begin{equation}
    P_0 = C \rho_0^{\gamma_Q},
    \label{eq:Pressure}
\end{equation}

where 
\begin{equation}
    \gamma_Q=\frac{a-b+\alpha-1}{\alpha-b}.
    \label{eq:Gamma_Q}
\end{equation}

For the stationary state to exist, not only the thermal balance but also the mechanical balance have to take place, which implies the equilibrium between gravitational force and the pressure gradient. Assuming the previously mentioned geometry, the mechanical equilibrium can be written as follows:
\begin{equation}
    \frac{dP_0}{dz} + \rho_0 g = 0.
	\label{eq:stationaryState1}
\end{equation}

Thus, we have the stationary state defined by thermal (\ref{eq:Pressure}) and mechanical (\ref{eq:stationaryState1}) balance conditions. Using these equations, we can now proceed to determine the height  distribution of plasma parameters. Let us assume that in the considered temperature range $\alpha=const_1$, and some heating mechanism is present in the medium such that $a=const_2, b=const_3$. Hence the index $\gamma_Q = const$ must be also constant.

In light of the foregoing, substitution of the pressure-density relation (\ref{eq:Pressure}) and subsequent integration of Eq.~(\ref{eq:stationaryState1}) with respect to the coordinate $z$ gives us the distribution of equilibrium pressure with height (\ref{eq:isothermicStationaryProfileP}). Then, using relation (\ref{eq:Pressure}) and equation of state (\ref{eq:EOS}), one can obtain the height distribution of density (\ref{eq:isothermicStationaryProfileRho}) and temperature (\ref{eq:isothermicStationaryProfileT}).
\begin{equation}
P_0(z)=P_{*} \left( 1 - \frac{z (\gamma_Q - 1)}{H \gamma_Q} \right)^{\frac{\gamma_Q}{(\gamma_Q-1)}},
	\label{eq:isothermicStationaryProfileP}
\end{equation}
\begin{equation}
\rho_0(z)=\rho_{*} \left( 1 - \frac{z (\gamma_Q - 1)}{H \gamma_Q} \right)^{\frac{1}{(\gamma_Q-1)}},
	\label{eq:isothermicStationaryProfileRho}
\end{equation}
\begin{equation}
T_0(z)=T_{*} \left( 1 - \frac{z (\gamma_Q - 1)}{H \gamma_Q} \right),
	\label{eq:isothermicStationaryProfileT}
\end{equation}
where   $H=\frac{P_{*}}{g \rho_{*}} = \frac{k_B T_{*}}{m g}$ is the characteristic gravitational spatial scale, $P_{*}$, $\rho_{*}$, $T_{*}$ are equilibrium pressure, density, and temperature at $z=0$, respectively. Within the framework of this work, one may choose as  level $z=0$ the height in the solar atmosphere for which the  parameters of plasma (temperature, pressure, and density) are specified (e.g using data from EUV or multi-frequency radio emission, the data from Interface Region Imaging Spectrograph (IRIS), or white light measurements.

It is clearly seen that considered plasma non-adiabaticity leads to the previously mentioned linear dependency of temperature with height (\ref{eq:isothermicStationaryProfileT}). Note that the slope of the line is completely defined by the heating/cooling rates and their dependence on temperature and density. Moreover, the heating/cooling processes affect the density and pressure profiles making them generally power-law rather than exponential. Thus, the observed non-exponential profile of density in the solar atmosphere can be attributed to non-adiabatic effect and furthermore can be used as a seismological tool to define possible corona heating mechanism.

In other words, the observed density profile can be fitted by function (\ref{eq:isothermicStationaryProfileRho}), finding the value of $\gamma_Q$. This, according to expression (\ref{eq:Gamma_Q}), allows us to determine the relationship between the parameters $a$ and $b$. This relation allows us to determine the range of possible values of $a$ and $b$. Such estimates may prompt an idea of what mechanisms may be responsible for the resulting heating dependence on temperature and density.

It is important to note that this approach will be viable if the stationary state defined by expressions (\ref{eq:isothermicStationaryProfileP}) -- (\ref{eq:isothermicStationaryProfileT}) is stable. This refers primarily to thermal instabilities. Thus, the conditions of isochoric ($(\partial Q/\partial T)_\rho > 0$) and isobaric ($(\partial Q/\partial T)_\rho - \rho_0 (\partial Q/\partial \rho)_T /T_0 > 0$) stability introduced by \citep{Field1965} must be fulfilled. These conditions of isochoric and isobaric stability for cooling (\ref{Cooling}) and heating (\ref{Heating}), written in the form of power functions of density and temperature take the following form, respectively:

\begin{equation}
\alpha -b > 0,
	\label{eq:isochoricStability}
\end{equation}

\begin{equation}
\frac{a-1}{\alpha-b} > -1.
	\label{eq:isobaricStability}
\end{equation}

Stability of the stationary state under conditions of isentropic instability ($(\partial Q/\partial T)_\rho + \rho_0 (\partial Q/\partial \rho)_T / (\gamma-1) T_0 < 0$) should be investigated in more detail. On the one hand, in a homogeneous medium, it does not affect the equilibrium state itself, although it leads to the emergence of quasi-periodic acoustic structures \citep{Zavershinskii2019}, which subsequently evolve into shock waves \citep{Molevich2021POF}. On the other hand, the isentropic instability condition implies a decrease in temperature height faster than the adiabatic gradient which is the instability condition for the gravity mode of AGWs \citep{brasseur_jacob_2017}. Under these conditions, one can expect the emergence of convective flows, which will disrupt the state of equilibrium. Therefore, we suppose that the condition of isentropic stability (\ref{eq:isentropicStability}), which is simultaneously a condition of stability of the gravity mode, should also be satisfied:

\begin{equation}
\frac{a-1}{\alpha-b} < \gamma-1.
	\label{eq:isentropicStability}
\end{equation}

The regions of thermal stability on $a-b$ diagram can be found in \cite{Kolotkov_2020}.

Another limitation of the approach described in this paper is related to the fact that the assumption of constancy of parameters $a$, $b$, $\alpha$ in heating (\ref{Heating}) and cooling (\ref{Cooling}) functions and, as a consequence, constancy of $\gamma_Q$ (\ref{eq:Gamma_Q}) in these formulas is used when deriving the height profiles of pressure, density, and temperature (\ref{eq:isothermicStationaryProfileP}) -- (\ref{eq:isothermicStationaryProfileT}). These assumptions may be applicable in some relatively small region, for example, so that the coefficient $\alpha$ in the radiative cooling function (\ref{Cooling}) is constant in this region. In addition, it is necessary to estimate the influence of thermal conduction in regions with temperature nonlinearities, since it will begin to influence the stationary state in Eq.~(\ref{eq:EnergyConservation}) in contrast to the linear height temperature profile.

Nevertheless, the power-law pressure (\ref{eq:isothermicStationaryProfileP}) and density (\ref{eq:isothermicStationaryProfileRho})  profiles can become exponential in the non-adiabatic plasma:
\begin{equation}
\rho_0(z)=\rho_{*} e^{-z/H}, P_0(z)=P_{*} e^{-z/H}.
	\label{eq:isothermicStationaryProfile}
\end{equation}
Such profiles imply temperature profile to be isothermal, which according to Eqs.~(\ref{eq:isothermicStationaryProfileP}) -- (\ref{eq:isothermicStationaryProfileT}) requires $\gamma_Q=1$. Furthermore, according to Eq.~(\ref{eq:Gamma_Q}), the parameter $a$ in the heating function (\ref{Heating}) have to be equal unity $(a=1)$ in the isothermal plasma. 
Thus, the quantity $b$  becomes the only free parameter in the heating function (\ref{Heating}) in this case. Note that if the atmosphere is isothermal and isochorically stable (\ref{eq:isochoricStability}), inequalities (\ref{eq:isobaricStability}), (\ref{eq:isentropicStability}) are always satisfied.

In what follows, we will consider how the thermal misbalance affect dispersion properties of waves under assumption of the isothermal temperature profile.

\section{Isothermal case}
\label{s:Isothermic case}
\subsection{Dispersion relation}
\label{ss:Dispersion relation}

Let us analyze  the wave properties in a gravitationally stratified isothermal atmosphere taking into account heating and cooling processes. To do this, we can apply the perturbation theory and seek a solution of Eqs.~(\ref{eq:MassConservation}) -- (\ref{eq:EOS}) using following substitution:
\begin{equation}
\xi(z,x,t) = \xi_0(z) (1 + \xi_1 (z,x,t)).
\label{linearization1}
\end{equation}

Hereinafter, the quantity $\xi$ means any parameter describing the state of the plasma. Subscript “0” means  the unperturbed plasma state. The quantity $\xi_1$ is for relative perturbation  solution. Assuming the  wave  propagating in the \textit{x}-\textit{z} plane (i.e. considering the wave-vector as  $\vect{k} = k_x\vect{e_x} + k_z\vect{e_z} $), we can use following  harmonic wave solution:
\begin{equation}
\xi_1(z,x,t) = \widetilde{\xi_{1}}\ exp \left\{ i k_x x + i k_z z - i \omega t \right\}.
\label{linearization2}
\end{equation}
where, $\widetilde{\xi_{1}}$ is the relative amplitude of the wave, and $\omega $ is the wave frequency.

In Eq.~(\ref{linearization1}), we 
adhere to the chosen geometry and consider the equilibrium parameters as a functions of the vertical coordinate only. Thus, we can define pressure and density gradients in  the isothermal atmosphere with $T_0 = {const}$ and $\rho_0(z)$, $P_0(z)$ determined by Eq.~(\ref{eq:isothermicStationaryProfile}) as follows:
\begin{equation}
\frac{d\rho_0}{dz} = -\frac{1}{H}\rho_0(z), \frac{d P_0}{dz} = -\frac{1}{H}P_0(z).
	\nonumber
\end{equation}

Non-adiabatic processes deserve special attention in the context of the linearization of Eqs.~(\ref{eq:MassConservation}) -- (\ref{eq:EOS}). Since heating and cooling processes depend on density and temperature, their influence on the propagating perturbations is determined by derivatives of function $Q(\rho,T)$ with respect to $\rho$ and $T$. In our analysis, we assume that for any equilibrium temperature $T_0(z)$
\begin{equation}
\frac{\partial Q}{\partial T} |_{T=T_0(z)} = const.
	\nonumber
\end{equation}

The derivative of $Q(\rho, T)$ with respect to $\rho$ can be found from the following considerations. In isothermal atmosphere, $\gamma_Q = 1$ that implies $a = 1$ in the cooling function. Hence, $Q (\rho, T) = \rho (\psi T^\alpha - h T^b)$. Since the derivative of the heat-loss function $Q$ is taken at equilibrium values of temperature $T = T_0$ and density $\rho = \rho_0$, by virtue of Eq.~(\ref{eq:Thermalbalance})
\begin{equation}
\frac{\partial Q}{\partial \rho} = 0.
	\nonumber
\end{equation}

Considering the mentioned above, linearization procedure applied to Eqs.~(\ref{eq:MassConservation}) -- (\ref{eq:EOS}) using substitution (\ref{linearization1}) -- (\ref{linearization2}) gives us the following set of equations
\begin{equation}
-i \omega \widetilde{\rho}_1- \frac{1}{H} \widetilde{V}_{z1} + i k_z \widetilde{V}_{z1} + i k_x\widetilde{ V}_{x1} = 0, \label{initLinear1}
\end{equation}
\begin{equation}
\omega \rho_0  \widetilde{V}_{x1} = k_x P_0 \widetilde{P}_1, \label{initLinear2}
\end{equation}
\begin{equation}
i \omega \widetilde{V}_{z1} = - \left( \frac{1}{H} - i k_z \right) H g \widetilde{P}_1 + \widetilde{\rho}_1 g, \label{initLinear3}
\end{equation}
\begin{equation}
i \omega \widetilde{T}_1 - (\gamma-1) \left( i \omega \widetilde{\rho}_1 + \frac{\widetilde{V}_{z1}}{H} \right) = - \frac{1}{\tau_V} \rho_0 T_0 \widetilde{T}_1, \label{initLinear4}
\end{equation}
\begin{equation}
\widetilde{P}_1 = \widetilde{\rho}_1 + \widetilde{T}_1. \label{initLinear5}
\end{equation}

Here, $\tau_V$ is the characteristic time associated with the thermal misbalance:
\begin{equation}
\tau_V = \frac{C_V}{\frac{\partial Q}{\partial T} |_{T=T_0(z)}}.
	\label{tauV}
\end{equation}

Simplification of Eqs.~(\ref{initLinear1}) -- (\ref{initLinear5}) gives us the dispersion relation in form:
\begin{eqnarray}
  \omega^2 - c_{S}^2 \left( k_x^2 + k_z^2 \right) - i \gamma g k_z + \frac{(\gamma-1) g^2 k_x^2}{\omega^2}  = \nonumber \\  = \frac{-i}{\omega \tau_V}\left(\omega^2 - c_T^2 \left( k_x^2 + k_z^2 \right) - i g k_z \right),
	\label{eq:dispersionRelationRaw}
\end{eqnarray}
where $c_T^2 = k_B T_0 / m$ is the square of isothermal sound speed and $c_S^2 = c_T^2 \gamma$ is the square of sound speed.

Let us transform Eq.~(\ref{eq:dispersionRelationRaw}) using substitutions similar to \cite{Priest2014}, namely $k_z^{'} = k_z + i/2H$. From geometry of problem, one can see that $k_x^2 + k_z^{'2} = k^{'2}$ and $k_x^2 = k^{'2} sin^2 \theta$, where $\theta$ is the angle between $k_z^{'}$ and $k^{'}$. Then, we can rewrite dispersion relation~(\ref{eq:dispersionRelation}) for AGW in the thermally active plasma as follows:
\begin{eqnarray}
\omega^2 - c_{S}^2 k^{'2} - N_S^2 + \frac{N^2}{\omega^2} c_S^2 k^{'2} sin^2  \theta =   \nonumber \\
=  \frac{-i}{\omega \tau_V}\left(\omega^2 - c_T^2 k^{'2} - N_{SQ}^2 \right)  = 0.
	\label{eq:dispersionRelation}
\end{eqnarray}

Here, $N_S = c_S/2H$  is the acoustic cut-off frequency and $N = g \sqrt{\gamma - 1}/c_S$  is the Brunt–Väisälä frequency. 

One may notice that the expression on the right-hand side of Eq.~(\ref{eq:dispersionRelation}) describes the contribution of thermal misbalance to the dispersion properties of AGW. Analysis of the derived equation reveals that the introduced parameter $N_{SQ}$ has the physical meaning of acoustic cut-off frequency in the medium with powerful processes of heating and cooling (strong thermal misbalance):
\begin{equation}
N_{SQ} = \frac{c_T}{2H} .
	\label{NSQ}
\end{equation}

It is also worth to mention that obtained dispersion equation (\ref{eq:dispersionRelation}) reduces to the well-known Eq.~(\ref{eq:dispersionRelationPriest}) for AGW \citep[see][]{Priest2014} in the case of thermal misbalance absence ($\tau_V \rightarrow \infty$). The last implies that the expression in parentheses on the right-hand side of Eq.~(\ref{eq:dispersionRelation}) tends to 0:
\begin{equation}
 \omega^2 - c_{S}^2 k^{'2} - N_S^2 + \frac{N^2}{\omega^2} c_S^2 k^{'2} sin^2 \theta = 0.
	\label{eq:dispersionRelationPriest}
\end{equation}

\subsection{The significance of thermal misbalance in the description of acoustic-gravity waves}
\label{ss:The significance of thermal misbalance}

\begin{figure}
	\includegraphics[width=\columnwidth]{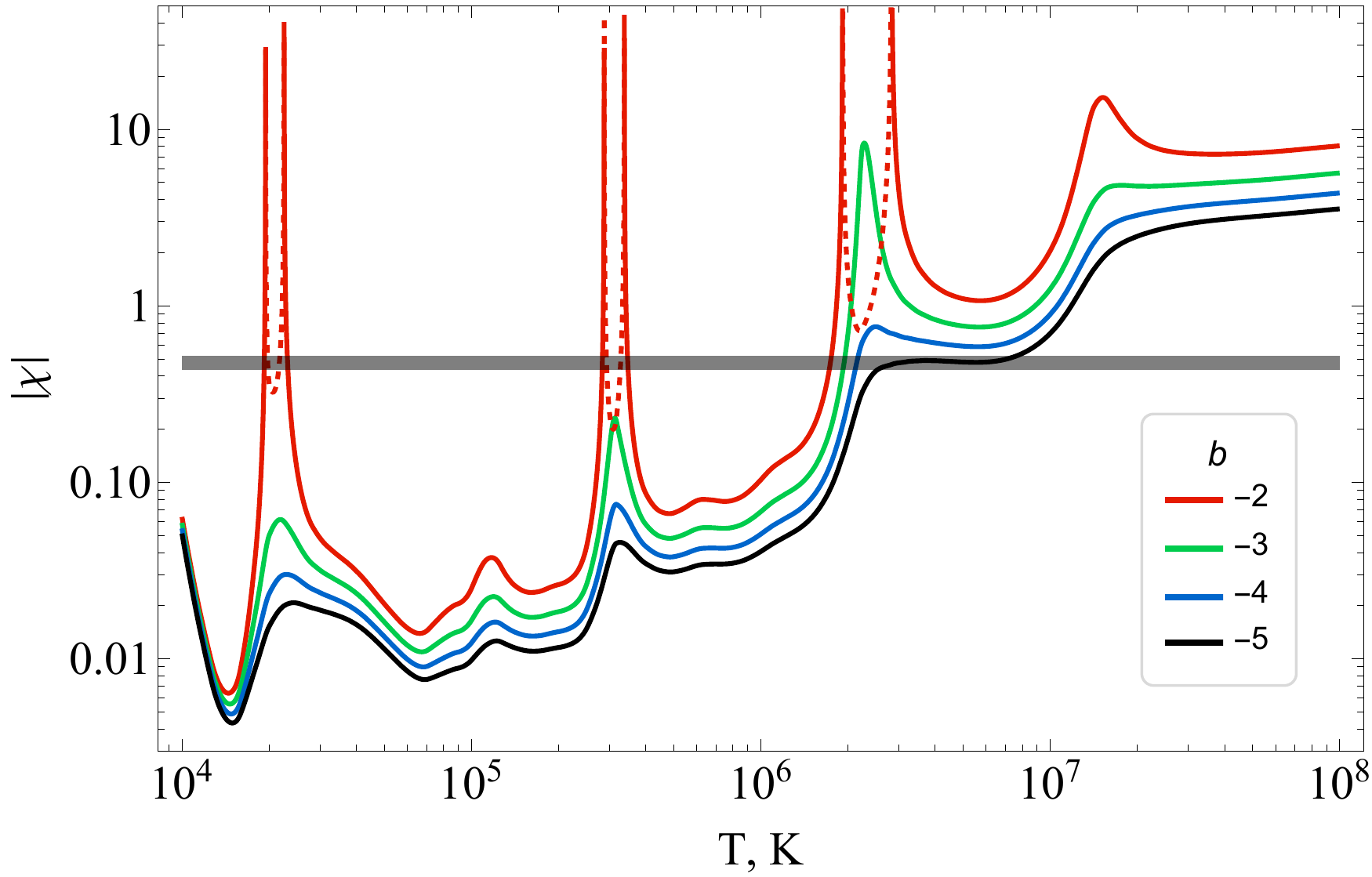}
    \caption{Dependence of the dimensionless parameter $\chi$ on temperature for several heating scenarios. The colored lines correspond to different form of the heating model (\ref{Heating}) with $a=1$ and $b$, given in the legend of the plot. The dashed lines indicate negative values of $\chi$. The gray line corresponds to the value $\chi$ below which gravitational waves become non-propagating over the entire spectrum (see Eq.~\ref{eq:GravityWavesNonPropagating}) for details).}  
    \label{fig:chi}
\end{figure}
\begin{figure}
	\includegraphics[width=\columnwidth]{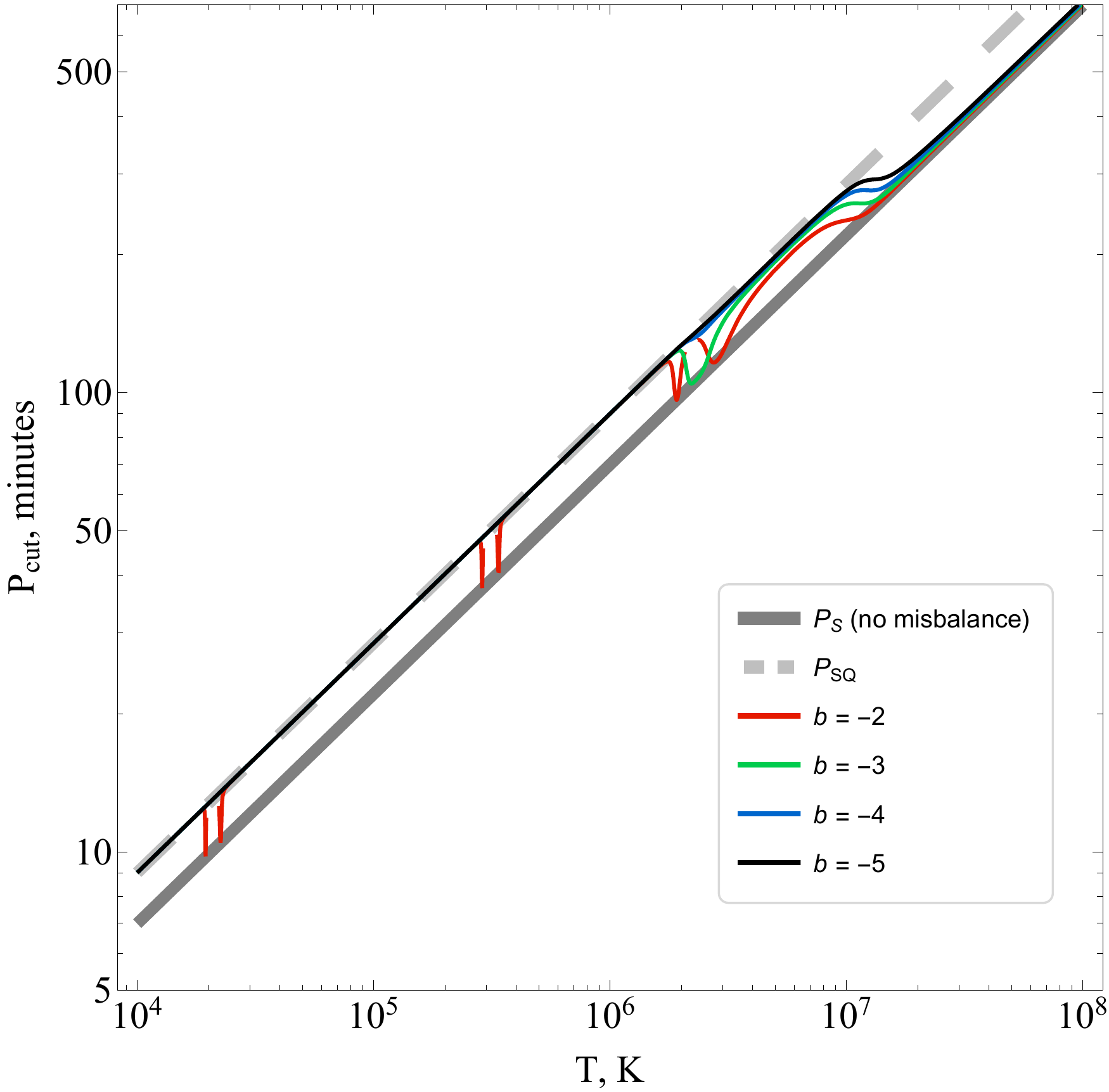}
    \caption{Dependence of acoustic cut-off period on temperature for several heating scenarios. The gray solid line indicate the cut-off period $P_S = 2 \pi / N_S$ in the plasma without thermal misbalance. The gray  dashed line corresponds to period  $P_{SQ} = 2 \pi / N_{SQ}$. The variations of the cut-off period with respect to the form of the heating model are shown by colored lines. }  
    \label{fig:Tcut-off}
\end{figure}
The analysis of Eq.~(\ref{eq:dispersionRelation}) shows that the dispersion properties of AGW significantly depends on dimensionless parameter $\chi$:
\begin{equation}
    \chi = \frac{c_T  \tau_V }{H}.
    \label{eq:chi}
\end{equation}

It can be interpreted as the ratio of characteristic length of thermal processes to the characteristic gravity scale.

The case  when $\chi \gg 1$ implies that the thermal misbalance affects waves on scales much larger than characteristic gravity scale $H$. In other words, the effect of thermal misbalance is relatively weak, and the wave dynamics is primarily determined  by gravity. This means that dispersion properties of AGW can be fairly accurately described by well-known  relation~(\ref{eq:dispersionRelationPriest}).

The opposite case when $\chi \ll 1$ implies that the thermal misbalance affects the waves at scales much smaller than gravity scale $H$. Thus, the effect of thermal misbalance will be clearly evident in the properties of the waves in this particular case, and the dispersion properties of AGW are determined by more general equation (\ref{eq:dispersionRelation}).

In the case when $\chi \approx 1$ the effects of gravity and thermal misbalance are of the same order.

Since this parameter $\chi$ is illustrative in determining the relative contribution of the effects of gravity and thermal misbalance, it is reasonable to estimate its value in  the solar atmosphere. Certainly, its value will depend on the specific type of heating and cooling functions. The latter is set in form~(\ref{Cooling}) with parametrization calculated using CHIANTI database \citep{Delzanna2020chianti}. The parameters in the heating function~(\ref{Heating}) are chosen so that the stationary state is isothermal and the medium is thermally stable over the vast temperature range. The first condition requires $a=1$ and the second condition is mostly satisfied with negative $b$.

The dependence of the parameter $\chi$ on temperature $T$ for several arbitrary values of $b$ in heating function~(\ref{Heating}) is plotted in Fig.~\ref{fig:chi}. One can see that $\chi \ll 1$ in the temperature range $T < 2~\mathrm{MK}$ and $\chi \sim 1$ for temperatures $2~\mathrm{MK} < T < 10~\mathrm{MK}$. It means that the effect of thermal misbalance on AGW is more important or at least as important as the effect of gravity for almost all feasible temperatures in the solar atmosphere. Their dispersion properties are therefore to be described by relation~(\ref{eq:dispersionRelation}) rather than by usual relation~(\ref{eq:dispersionRelationPriest}).

Note that in some temperature ranges shown in Fig.~\ref{fig:chi}, the $\chi$ values are negative (see dashed lines). This is due to the fact that in these cases, the thermal misbalance will lead to the amplification of compressional waves. Therefore, these temperature ranges are beyond the scope of our consideration.

Analysis of dispersion relation~(\ref{eq:dispersionRelation}) also shows that thermal misbalance leads to a decrease in the acoustic cut-off frequency (increase in the cut-off period) at $\chi \lesssim 1$. Fig.~\ref{fig:Tcut-off} shows that this increase is noted in different heating scenarios up to a temperature of 10 MK, which also indicates the importance of thermal misbalance in the dynamics of acoustic waves over a large temperature range.

\subsection{Dispersion properties of acoustic-gravity waves in a medium with thermal misbalance}
\label{ss:Dispersion properties of acoustic-gravity}

\subsubsection{Acoustic mode/Slow mode}
\label{sss:Acoustic mode}

\begin{figure*}
	\includegraphics[width=\textwidth]{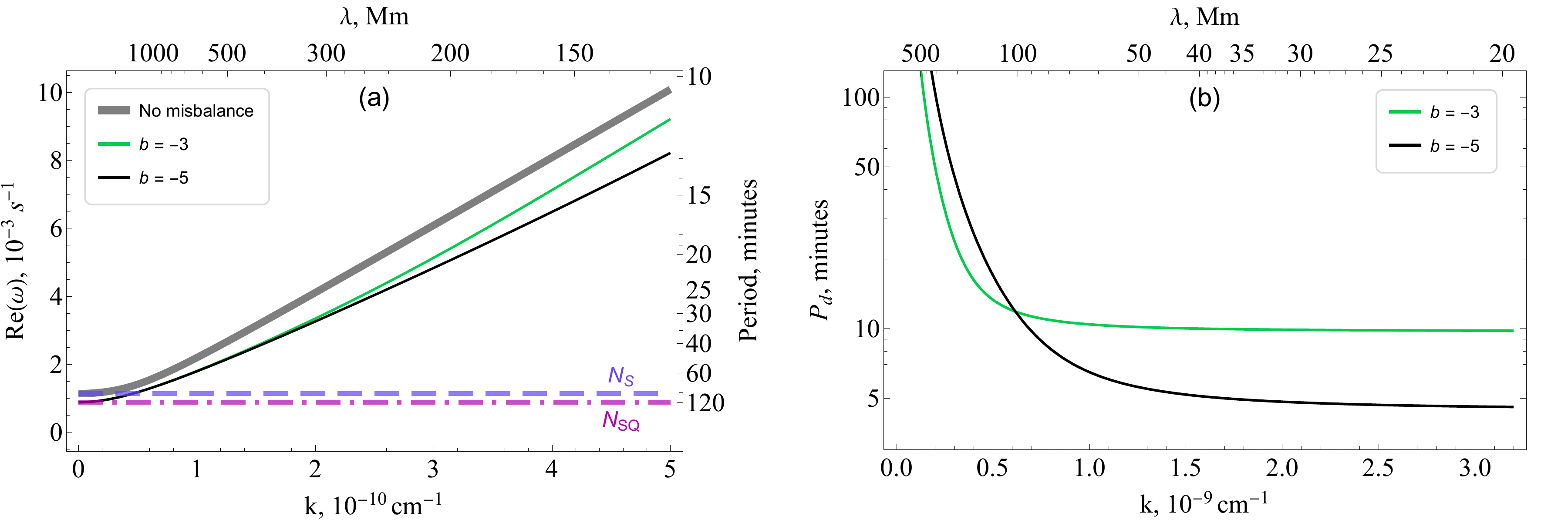}
    \caption{Dependence of (a) real part of frequency $\omega$ and (b) characteristic damping time of acoustic waves on wavenumber $k$ in an isothermal plasma with thermal misbalance under solar coronal conditions.  Calculations are made for temperature $T = 1.75~\mathrm{MK}$ and mean molecular weight $\mu=0.6$.}
    \label{fig:ReIm}
\end{figure*}

\begin{figure*}
	\includegraphics[width=\textwidth]{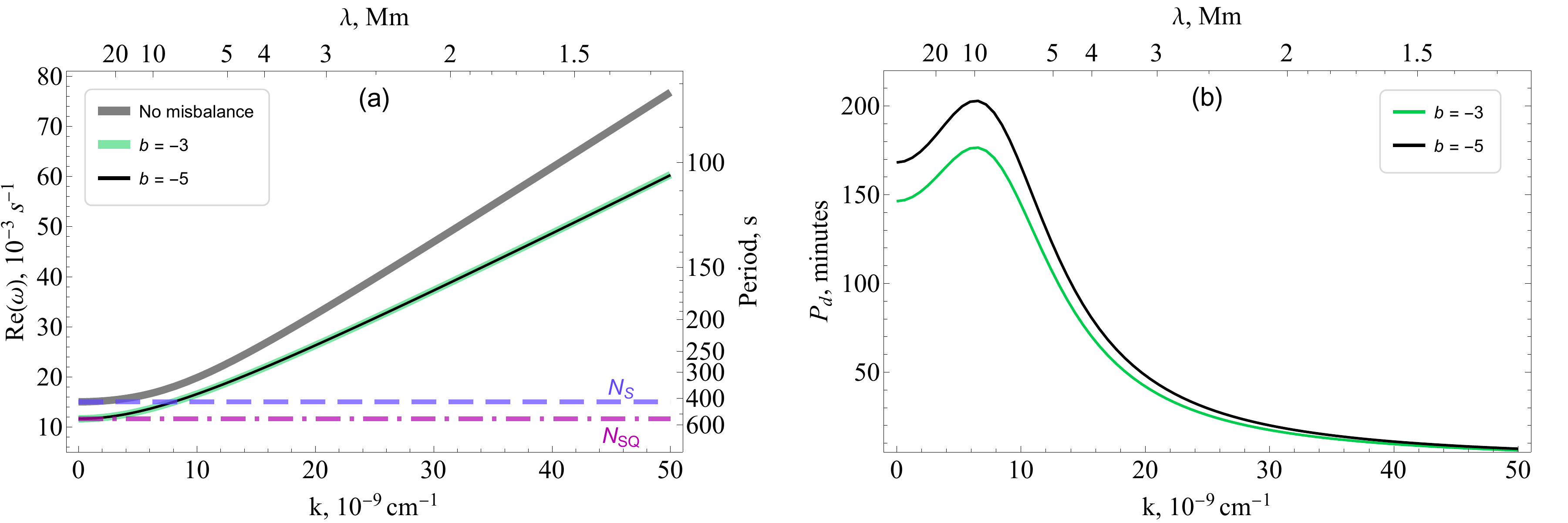}
    \caption{Dependence of (a) real part of frequency $\omega$ and (b) characteristic damping time of acoustic waves on wavenumber $k$ in an isothermal plasma with thermal misbalance under chromospheric conditions. Calculations are made for temperature $T=10,000K$ and mean molecular weight $\mu=0.6$.
    }
    \label{fig:acoustic_chromosphere}
\end{figure*}

\begin{figure}
	\includegraphics[width=\columnwidth]{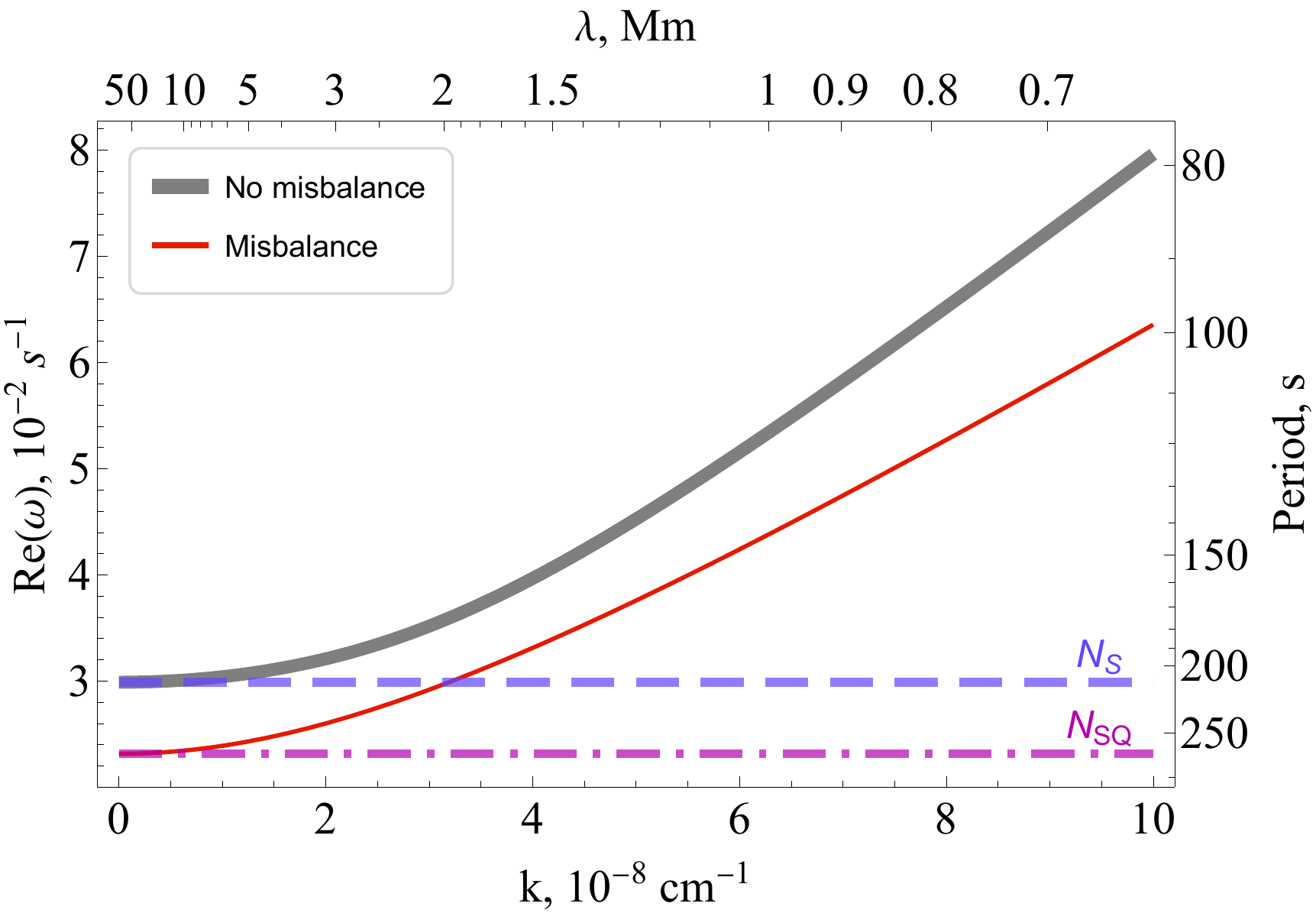}
    \caption{Estimated change in the dependence of real part of frequency $Re(\omega)$ on the wave number $k$ and cut-off frequency under the influence of thermal misbalance in photosphere assuming an isothermal atmosphere and powerful heating and cooling processes ($\chi \ll 1$).
    }
    \label{fig:acoustic_photosphere}
\end{figure}

Dispersion relation~(\ref{eq:dispersionRelation}) is a 4th degree equation with respect to $\omega$, so it can be solved analytically. However, this solution, given the complex coefficients, is difficult to analyze. So, we illustrate this solution for acoustic modes for different regions of the solar atmosphere.

In our calculations for slow modes in the solar corona, we use typical equilibrium temperature  $T_0 = 1.75~\mathrm{MK}$ from \cite{roberts_2019} and the mean molecular weight $\mu = m/m_H=0.6$. Calculation of heating and cooling is done in the same way as in the previous paragraph. The calculation of real part of frequency $\omega$ and characteristic damping time ($P_d = 1/Im(\omega)$) over real wavenumber $k$ for solar corona conditions is shown in Fig.~\ref{fig:ReIm}.

One can see from Fig.~\ref{fig:ReIm} that the graphs corresponding to the effect of thermal misbalance lie below the graph without it. This always occurs in the isothermal atmosphere (i.e., when $\gamma_Q = 1$). Analysis of dispersion relation~(\ref{eq:dispersionRelation}) shows that acoustic cut-off frequency can decrease down to $N_{SQ}$ at $\chi \ll 1$ that is $\sqrt{\gamma}$ times smaller than traditional cut-off frequency $N_S$ in the medium without thermal misbalance. At $\chi \sim 1$, the acoustic cut-off frequency is between $N_S$ and $N_{SQ}$. Our estimations for different heating scenarios show that the thermal misbalance is sufficiently powerful to decrease acoustic cut-off frequency (increase cut-off period) by $\sqrt{\gamma}$ in almost all temperature ranges up to 10 MK (see Fig.~\ref{fig:Tcut-off}).

Also, note that the slope angle of different graphs corresponding to the case of thermal misbalance may differ for certain wavelength regions (Fig.~\ref{fig:ReIm}a). This is the result of dispersion of phase speed caused by both thermal misbalance and gravity.

Another effect of thermal misbalance is an additional damping of slow waves \citep{2019A&A...628A.133K}. Our calculations for different heating mechanisms show a characteristic damping time of slow waves of 5-10 minutes (Fig.~\ref{fig:ReIm}b). That is, the damping time coincides in order of magnitude with the characteristic periods of the observed waves \citep{2019ApJ...874L...1N}.

Summarizing, one can say that thermal misbalance is indeed important effect for wave dynamics in solar corona and chromosphere to take into account. However, due to the high temperature of the solar corona and, consequently, the very large gravitational height $H$, the characteristic scales at which one must account for gravity and thermal imbalance simultaneously are beyond the size of the medium. For most applications in the solar corona, it is sufficient to consider thermal misbalance only.

In the photosphere, on the contrary, such consideration may be necessary due to its low temperature and relatively small gravitational height $H$. The description of heating and cooling functions in form~(\ref{Heating}), (\ref{Cooling}) can be applied to the photospheric/chromospheric conditions using some strong assumptions \citep{2022SoPh..297..144I, 1986ApJ...308..975A}, since the optically thin radiation becomes less dominant. Then, based on the analytical results obtained above and on the extension of the radiative cooling function, we can conclude that the regime with relatively powerful heating and cooling processes ($\chi \leq 1$) may take place in the photosphere. Estimations made with our simple model for the isothermal atmosphere with photospheric conditions (with temperature $T_0 = 5350 \mathrm{K}$ and mean molecular weight $\mu=1.27$) show the increase of the period of waves above which the waves become non-propagating from 210 s up to 270 s in the $\chi \ll 1$ regime (Fig.~\ref{fig:acoustic_photosphere}). Nevertheless, this result primarily means that it makes sense to consider the influence of thermal processes on the properties of waves in the photosphere with a more appropriate model, taking into account the radiative transfer equation.

\subsubsection{Gravity mode}
\label{sss:Gravity mode}

\begin{figure}
	\includegraphics[width=\columnwidth]{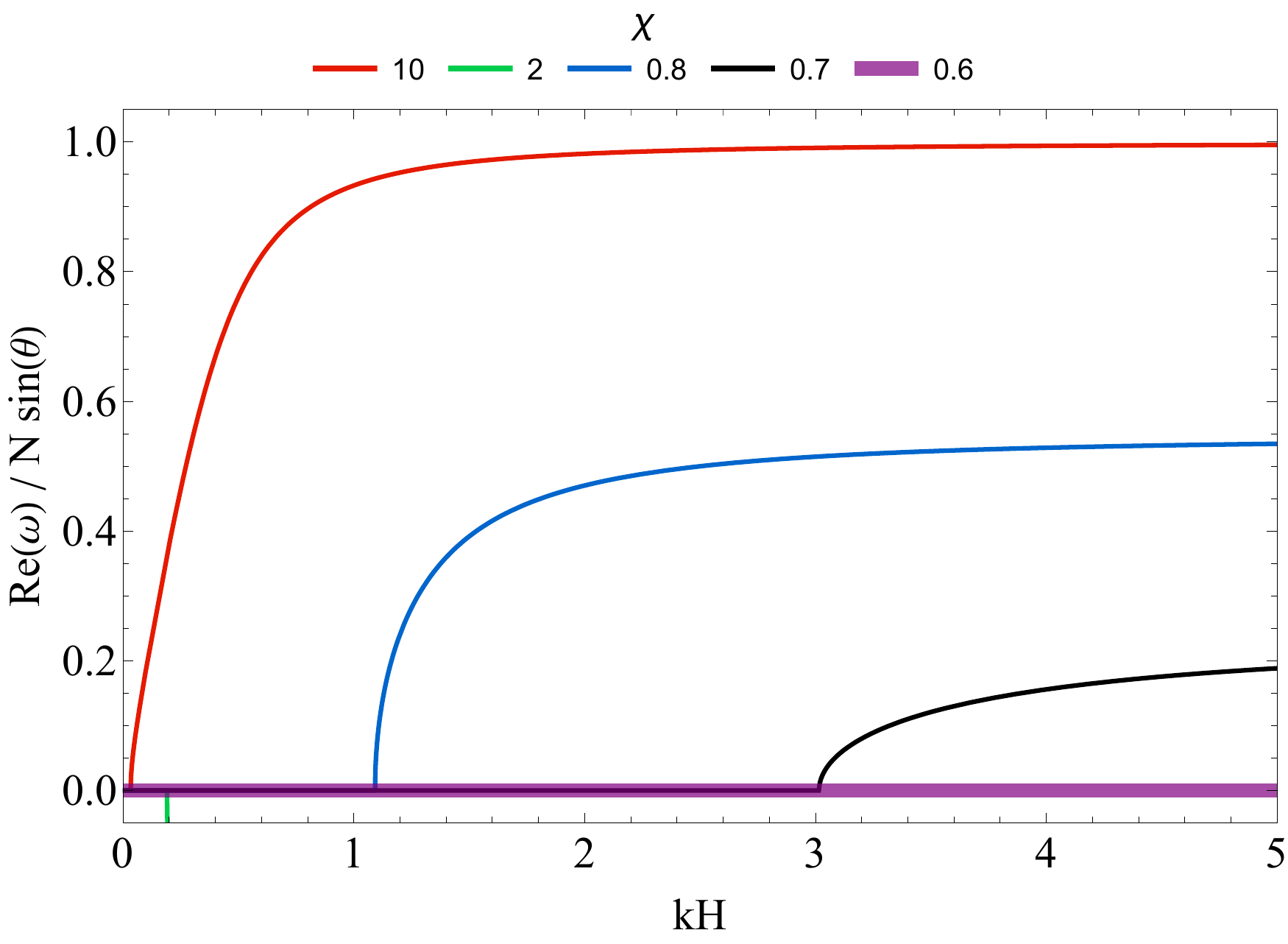}
    \caption{Dependence of the real part of the frequency of the gravity waves on the dimensionless wave number at different $\chi$.}
    \label{fig:GravityWaves}
\end{figure}

Analysis of dispersion relation~(\ref{eq:dispersionRelation}) shows that thermal misbalance also significantly affects the gravity waves. It is a mode primarily driven by buoyancy force. The mechanism of this force is illustrated by the oscillation of the blob of plasma near an equilibrium state. The blob is displaced a distance $\delta s$ (vertical displacement is $\delta z = \delta s \cos \theta$). As a result of changes in the background thermodynamic parameters, the density and pressure in the blob itself also change to be in a state of pressure equilibrium with its surroundings. Such considerations lead to the dispersion relation for internal gravity waves $\omega^2=N^2 \sin^2 \theta$. Compressibility and buoyancy, acting together, slightly modify the properties of the gravity mode but in the limit $\omega \ll k^{'} c_s$ their dispersion relation is reduced to the dispersion relation for internal gravity waves. The account of the thermal misbalance should lead to a change in the blob oscillation frequency because the change in density and temperature when it is displaced along $\delta s$ will cause a change in the heating and cooling power, which in turn will lead to a change in the thermodynamic parameters of this blob. The study of the dispersion relation~(\ref{eq:dispersionRelation}) at $k \rightarrow \infty$ shows that the frequency of oscillations of gravity waves will be determined by the following expression:

\begin{equation}
    \omega_{osc} = \sqrt{N^2 sin^2{\theta}-\frac{1}{4\tau_V^2\gamma^2}} = N \sqrt{sin^2{\theta}-\frac{1}{4 \chi^2 \gamma (\gamma-1)}}.
    \label{eq:BruntVaisalaMisbalance}
\end{equation}

The obtained relation can be interpreted as an analog of the  Brunt–Väisälä frequency in a plasma with thermal misbalance. If the effect of thermal misbalance in relation to gravity is sufficiently weak ($\tau_V, \chi  \rightarrow \infty $), then frequency $\omega_{osc}$ reduces to $N \sin{\theta}$, which coincides with that for plasma without thermal misbalance. The decrease of $\chi$ corresponding to the increase of the thermal misbalance impact in relation to gravity leads to the decrease of $\omega_{osc}$ (see colored lines in Fig.~\ref{fig:GravityWaves}) down to 0 (see purple line in Fig.~\ref{fig:GravityWaves})). 

Moreover, the thermal misbalance can not only reduce the frequency of oscillations of gravity waves but also make it imaginary. It means that gravity waves can become non-propagating (see Fig.~\ref{fig:GravityWaves}) in a medium with sufficiently powerful processes of heating and cooling (low $ \chi $). One can derive from Eq.~(\ref{eq:dispersionRelation}) that gravity waves become non-propagating at any angles at
\begin{equation}
    \chi \leq \frac{1}{2 \sqrt{\gamma (\gamma-1)}}.
    \label{eq:GravityWavesNonPropagating}
\end{equation}

For coronal plasma with $\gamma=5/3$, this condition transforms to  $\chi \lesssim 0.47$ (represented by the gray line in Fig.~\ref{fig:chi}). One can see from Fig.~\ref{fig:chi} that in isothermal plasma gravity waves behave as non-propagating damped perturbations at temperatures below 2 MK.

\section{Discussion and conclusions}
\label{s:Discussion}

In the current research, we have investigated the joint effect of gravity and thermal misbalance on the plasma state and on the dynamics of AGW analytically. We use the well-known power-law model of the optically thin radiation~(\ref{Cooling}) \citep[e.g.,][]{Delzanna2020chianti} and assume that the heating mechanism is some general power-law function (\ref{Heating}) as well. Such an assumption makes it possible to obtain a fairly large set of results and implications that may be of interest in the context of coronal and MHD-seismology.

To begin with, let us discuss the results concerning the equilibrium state of the plasma. It follows from our analysis that heating (\ref{Heating}) and cooling (\ref{Cooling}) given in the form of power-law functions  imply generally linear dependency of temperature with height (see Eq.~(\ref{eq:isothermicStationaryProfileT})). Moreover, the non-adiabatic processes affect not only temperature but also the density and pressure profiles making them generally power-law rather than exponential (see Eqs.~(\ref{eq:isothermicStationaryProfileP}), (\ref{eq:isothermicStationaryProfileRho})). It is important to note that the slope of the density, temperature, and pressure profiles are completely determined by the form of the radiation loss and heating functions. In Eqs.~(\ref{eq:isothermicStationaryProfileP}) - (\ref{eq:isothermicStationaryProfileT}), the contribution of these non-adiabatic functions is represented using the exponent $\gamma_Q$ (\ref{eq:Gamma_Q}). The relation between the non-adiabatic processes and height profiles introduces an additional approach to obtaining the seismological constraints on the coronal heating mechanism.

Schematic procedure is as follows. Firstly, one has to define the height profiles of the coronal plasma. The height density and temperature profiles of the solar corona can be obtained using different techniques including forward modeling EUV Emission Observed by SDO/AIA \citep[see][for details]{2019ApJ...884...43P}, the Interface Region Imaging Spectrograph (IRIS) \citep{2018ApJ...864...21K}, multi-frequency radio emission \citep{2015A&A...583A.101M} or white light measurements \citep{1999ApJ...510L..63E}. In general, the height profiles may vary from exponential form.  In this case, the non-exponential profile can be fitted by the power-law expressions (\ref{eq:isothermicStationaryProfileP}) -- (\ref{eq:isothermicStationaryProfileT}) allowing to specify the exponent $\gamma_Q$ (\ref{eq:Gamma_Q}). The latter is defined by the cooling rate exponent $\alpha$, and heating rate exponents, namely, $a$ and $b$. Assuming the radiation losses to be a verified function \citep[see e.g.][]{Delzanna2020chianti}, we consider $\alpha$ as a known parameter. Thus, using the fitted values of $\gamma_Q$, we obtain the equation, which allows us to determine the constraints on the parameters $a$, $b$ of the heating function (\ref{Heating}). The results of such application is of interest in the context of comparison with the constraints on the coronal heating model (parameters $a$, $b$) obtained using slow wave observations \citep{Kolotkov_2020,2022MNRAS.514L..51K, 2023Physi...5..193K}.

The non-adiabatic processes affect not only the plasma state but also the compressional perturbations, namely, AGW, modifying their dispersion properties. The dispersion relation for AGW obtained using the assumption of isothermal atmosphere has the form of Eq.~(\ref{eq:dispersionRelation}). Here, we consider two main sources of dispersion, which are thermal misbalance and gravitational stratification. The characteristic temporal/spatial scales, where the dispersion effect is most pronounced, are different for these processes. In order to define for which temperature range a given effect will be the primary one, we introduced the dimensionless ratio $\chi$ (\ref{eq:chi}). Using the parametrization of the radiation loss function (\ref{Cooling}) calculated on the basis of CHIANTI database \citep{Delzanna2020chianti}, we obtain the dependence of $\chi$ on temperature for several heating scenarios (see Fig.~\ref{fig:chi}). The heating scenarios have been chosen in such a way that compressional waves are stable (decaying) over the vast temperature range considered. Stable state corresponds to $Im (\omega) < 0$ in dispersion relation (\ref{eq:dispersionRelation}) (solid lines in Fig.~\ref{fig:chi}). It follows from the obtained plot that the thermal misbalance is the primary dispersion source for AGW  for temperatures $T < 2~\mathrm{MK}$ and comparable to gravity effect for the range $2~\mathrm{MK} < T < 10~\mathrm{MK}$. Thus, one may conclude that the consideration of thermal misbalance is quite important for the problems of wave propagation in the stratified solar atmosphere.

Next, let us discuss how the non-adiabatic processes affect the properties of acoustic waves in the stratified isothermal plasma. Our calculations reveal that the thermal misbalance increases the periods of waves, which can propagate in the solar atmosphere (see Figs.~\ref{fig:ReIm}a, and  ~\ref{fig:acoustic_photosphere}). For considered coronal plasma temperature $T_0 = 1.75~\mathrm{MK}$, the variations in periods are $10 - 30~\%$ depending on the operating heating mechanism and considered wavenumber (see Fig.~\ref{fig:ReIm}a). The greatest variations concern longer waves. A similar effect can also take place for photospheric conditions (see Fig.~\ref{fig:acoustic_photosphere}).
However, the last conclusion is of limited applicability. Although photospheric plasma can be assumed to be ideal \citep{roberts_2019} and optically thin \citep{2013MNRAS.429.3133L}, these are still quite rough assumptions.  

Also, the thermal misbalance expectedly introduces damping (or amplification) in the gravitationally stratified plasma, as in the uniform \citep{2019A&A...628A.133K} and magnetically structured plasma \citep{2022MNRAS.514.5941A}. As we mentioned previously, in this paper we have focused on the thermal misbalance regime ($\chi \ll 1$), which implies the attenuation of compression waves over a wide range of temperatures. The damping periods of acoustic waves calculated for coronal conditions are shown in  Fig.~\ref{fig:ReIm}b. The obtained periods are of 5-10 minutes and coincide in order of magnitude with the characteristic periods of the observed waves \citep{2019ApJ...874L...1N}. It should be noted that in the case of positive values of the power index "$b$" in the heating model (\ref{Heating}), the regimes with instability of entropy and acoustic mode can take place. The acoustic instability regime allows formation of QPPs \citep{Zavershinskii2019} or even autowave shock pulses \citep{2020PhRvE.101d3204Z, 2010PhPl...17c2107C, Molevich2021POF}. The last issues are of interest in the context of formations of density perturbation by standing shock at acoustic cut-off period \citep{2017A&A...601A..42P}.

Speaking of wave/oscillation observations, the influence of thermal misbalance on the acoustic cut-off period merits special attention. Our analysis reveals that for the solar atmosphere cooler than $T < 2~\mathrm{MK}$, the assumed cut-off period should be increased by $\sqrt{\gamma}$ times (about $30 \%$). This issue is of interest of $\sim 90~\mathrm{min} $ oscillation indicated by \cite{2010SoPh..267..175V,2015ApJ...807..176V, 2017A&A...601A..42P}. The estimated plasma temperature $T \sim 1~\mathrm{MK}$ implies the cut-off period $P_{cut} \sim 70~\mathrm{min}$ calculated assuming the usual expression for the plasma without thermal misbalance. In turn, the introduced increased period value $N_{SQ}$ (\ref{NSQ}) gives observed $\sim 90~\mathrm{min} $ (see Fig.~\ref{fig:Tcut-off}). It should be noted that generally, the cut-off period lies between the  $N_{SQ}$ (\ref{NSQ}) and the usual value for plasma without thermal misbalance $N_S$   (see Fig.~\ref{fig:Tcut-off} for details). 

We obtain that non-adiabatic processes also significantly affect the dynamics of gravity waves. In particular, for the considered regime with stability/decaying of entropy and acoustic waves ($Im(\omega) < 0$), the thermal misbalance leads to the decay of gravity waves. However, as well as acoustic waves, gravity waves may become unstable ($Im(\omega) > 0$) for some heating mechanisms, which remained beyond our consideration (parameters $a$, $b$ in Eq.~(\ref{Heating}) affect dispersion properties through parameters (\ref{eq:Gamma_Q}), (\ref{tauV}) in dispersion relation (\ref{eq:dispersionRelation})).

In addition, the thermal misbalance in a stable isothermal atmosphere reduces the maximum oscillation frequency (analog of Brunt–Väisälä frequency) of gravity waves. One may notice from results shown in Fig.~\ref{fig:GravityWaves} that the increase of role of non-adiabatic processes (decrease of $\chi$ (\ref{eq:chi})) leads to the decrease of maximum oscillation frequency.

And the last but not least, the gravity waves can become non-propagating waves in the thermally active plasma. The above mentioned decrease of $\chi$ (\ref{eq:chi}) leads to the spectrum limitation  (see Fig.~\ref{fig:GravityWaves}). We show that the gravity waves become non-propagating at any angles if $\chi$ satisfy condition  (\ref{eq:GravityWavesNonPropagating}), which  takes the form $\chi \lesssim 0.47$ for coronal conditions. According to the results shown in Fig.~\ref{fig:chi}, such condition is satisfied for temperatures $T_0 \lesssim 1~\mathrm{MK}$. This result indicates that analysis of propagating gravity waves in non-adiabatic plasma requires further investigation and, in particular, under the conditions of non-isothermal height profile. 

The results presented in this paper are obtained using some assumptions. First, the medium is assumed to be fully ionized and the radiation is optically thin. This limits the applicability of the model to temperatures above about $10^4~\mathrm{K}$. For regions of the solar atmosphere with lower temperatures, the considered model should be applied with caution and for qualitative rather than quantitative description. A quantitative description requires a more complete model that takes into account partial ionization and radiation transport. Second, the height profile is assumed to be stable, and there are no large-scale flows. For this, at least the medium must be thermally stable, i.e., inequalities (\ref{eq:isochoricStability}) -- (\ref{eq:isentropicStability}) must be satisfied. For a more accurate consideration it is also required to further investigate the effects of the magnetic field and thermal conduction. In particular, thermal conduction will have a large effect on the steady state at large nonlinearities of the temperature profile. Additional corrections will be made by the dependence of the thermal conduction coefficient on temperature, which is important in an atmosphere with a relatively large temperature gradient, as well as the dependence of gravitational acceleration on height, which takes place in the solar corona because of its large size on the order of several solar radii.

As a final matter, it may be noted that the theory introduced in the current work can find its application not only for the analysis of waves in the atmosphere of the Sun but also in the stellar atmospheres \citep[see, e.g.,][]{2022ApJ...931...63L}.

\section*{Acknowledgements}

The study was supported in part by the Ministry of Education and Science of Russia by State assignment to educational and research institutions under Projects No. FSSS-2023-0009 and No. 0023-2019-0003.

CHIANTI is a collaborative project involving George Mason University, the University of Michigan (USA), University of Cambridge (UK), and NASA Goddard Space Flight Center (USA).

\section*{Data Availability}

The data underlying this article will be shared on reasonable request to the corresponding author.



\bibliographystyle{mnras}
\bibliography{refs} 







\bsp	
\label{lastpage}
\end{document}